\documentclass[twocolumn,showpacs,preprintnumbers,amsmath,amssymb]{revtex4}

\usepackage{amsmath}
\usepackage{graphicx}
\usepackage{epsf}
\usepackage{psfrag}
\usepackage{epsfig}
\usepackage{graphics}

\begin{document}

\newcommand{\be}{\begin{equation}}
\newcommand{\ee}{\end{equation}}
\newcommand{\bq}{\begin{eqnarray}}
\newcommand{\eq}{\end{eqnarray}}
\newcommand{\dt}{\frac{d^3k}{(2 \pi)^3}}
\newcommand{\dtp}{\frac{d^3p}{(2 \pi)^3}}

\title{{\bf{Implicit regularization beyond one loop order: scalar field 
theories}}}

\date{\today}

\author{Carlos R. Pontes$^{(a)}$} \email []{crpontes@fisica.ufmg.br}
\author{A. P. Ba\^eta Scarpelli$^{(b)}$} \email[]{scarp@fisica.ufmg.br}
\author{Marcos Sampaio$^{(c)}$}\email[]{marcos.sampaio@th.u-psud.fr}
\author{M. C. Nemes}\email[]{carolina@fisica.ufmg.br}

\affiliation{(a) Federal University of Minas Gerais -
Physics Department - ICEx \\ P.O. BOX 702, 30.161-970, Belo Horizonte
MG - Brazil}
\affiliation{(b) Centro Federal de Educa\c{c}\~ao Tecnol\'ogica - MG \\
Avenida Amazonas, 7675 - 30510-000 - Nova Gameleira - Belo Horizonte 
-MG - Brazil}
\affiliation{(c) Universit\'e Paris XI - Laboratoire de Physique 
Th\'eorique \\
B\^atiment 210 \,\,  F-91405 \,  Orsay \, Cedex. }

\begin{abstract}
\noindent Implicit regularization (IR) has been shown as an useful
momentum space tool for perturbative calculations in dimension
specific theories, such as chiral gauge, topological and
supersymmetric quantum field theoretical models at one loop level.
In this paper, we aim at generalizing systematically  IR to be
applicable beyond one loop order. We use a scalar field theory as an
example and  pave the way for the extension to quantum field
theories which are richer from  the symmetry content viewpoint.
Particularly, we show that a natural (minimal) renormalization
scheme emerges, in which the infinities displayed in terms of
integrals in one internal momentum are subtracted, whereas infrared
and ultraviolet modes do not mix and therefore leave no room for
ambiguities. A systematic cancelation of the infrared divergences at
any loop order takes place between the ultraviolet finite and
divergent parts of the amplitude for non-exceptional momenta
leaving, as a byproduct, a renormalization group scale.
\end{abstract}

\pacs{11.10.Gh, 11.15.Bt, 12.38.Bx}

\maketitle

\section{Introduction}
\indent

The problem of defining a  consistent perturbative regularization of
dimension specific quantum field theories is by itself of
theoretical interest. Chiral, topological and supersymmetric  gauge
theories belong to this class of theories. The very concept of
chirality is highly dimension dependent as there are no Weyl
fermions in odd dimensions. The $\gamma_5$ matrix which enters the
chiral projection operators has no straightforward generalization to
arbitrary dimensions. Hence, care must be exercised in applying
dimensional regularization methods. Yet the standard model as a full
theory is actually anomaly free, from the practical standpoint
perturbative calculations  are necessary to test the standard  model
and extensions against the experiments. The appearance of spurious
anomalies, which spoils renormalizability by violating chiral
symmetry, demands the addition of symmetry restoring counterterms,
order by order, in perturbation theory. This renders practical
calculations very tedious. Besides dimensional regularization (DREG)
explicitly breaks supersymmetry because the number of degrees of
freedom of gauge bosons and gauginos does not match for $d \ne 4$.
Therefore, an invariant regularization scheme becomes important in
this case. An alternative approach based on algebraic
renormalization, which discards an invariant regularization scheme,
is based on the construction of Slavnov-Taylor identities which
encodes all the essential symmetries of the model in consideration,
gauge and supersymmetry included. Such constraints, in turn, deliver
unambiguously the (non-invariant) counterterms to be added in order
to heal the symmetry breakdowns. The main drawback of this approach
is that it is much simpler to work within an invariant scheme which
hitherto is missing at least to all orders in perturbation theory.
Moreover, it is not a priori obvious that, in general, there are no
supersymmetry anomalies  \cite{STOCKINGER},  although the evidences quoted so
far points to supersymmetry being a full symmetry of the quantum
theory \cite{JACK-JONES}.

From the phenomenological viewpoint, it is expected that the minimal
supersymmetric standard model (MSSM)  can be realized at weak scale
in the experiments at LHC around $1 \,  TeV$. This makes quantum
effects important. That is to say, besides the direct detection of
susy particles, supersymmetry can also be probed via the virtual
effects of additional particles such as the eletroweak precision
observables within the MSSM \cite{HEINEMEYER}. Examples are the correction
for the $W$ boson mass, the effective leptonic weak mixing angle,
the anomalous magnetic moment of the muon. A  sound perturbative
calculation from the technical standpoint may also shed some light
on the supersymmetry breaking mechanism  by constraining the number
of parameters of the underlying soft breaking model when contrasted
with experiments .

Dimensional reduction (DRED), a framework where only momenta are
treated in $D$ dimensions, whereas $\gamma$ matrices and gauge
fields remain ordinary $4-$vectors, is used in most practical cases
where  the supersymmetry breaking of DREG matters, even though it
has been proved  mathematically inconsistent \cite{SIEGEL}. Nonetheless,
recent works have demonstrated that its validity can be extended
beyond one loop order by using the quantum action principle, although it remains to be circumvented some problems
related to QCD mass factorization \cite{STOCKINGER}. From the pragmatical
point of view, DRED is largely used to unravel supersymmetric
extensions of the standard model \cite{SPA}. However it is important to note
that  in DRED it is unclear whether or to what extent susy is
actually preserved. It is only known that some susy relations are
satisfied at the one and two loop level but even at one loop the
checks do not exhaust all the Green's functions that could be
affected by a susy breaking.

An important coordinate space regularization independent  framework,
which operates in the physical dimension of the model, is
Differential Renormalization (DR). DR basically consists of
replacing coordinate-space amplitudes that are too singular  to have
a Fourier tranform by derivatives of less singular ones as well as
an integration by parts prescription. It has been proved to be a
symmetry preserving scheme \cite{DIFF}. This is achieved by
both constraining the number of  independent scales which result
from each divergence and applying a  set of operational rules in the
introduction of basic functions (Constrained-DR), which assume the
same values as the original divergent ones except for coincident
points (short distance behaviour). Whilst at one loop order
Constrained-DR automatically produces expressions with a single mass
scale which fulfill the corresponding Ward-Slavnov-Taylor identities
of the underlying model, at two loops a consistent extension of this
program has not yet been constructed \cite{SEIJAS}.

Implicit regularization (IR) is a momentum space regularization
independent framework,  which, like Differential Renormalization,
does not rely on dimensional extension of space-time dimension
\cite{IR1}-\cite{IR12}. The basic idea behind IR is to assume
implicitly the presence of a regularization as part of divergent
amplitudes only in order to separate their regularization dependent
from its finite part, by using a simple algebraic identity in the
integrand of the amplitude. This operation is  analogous to apply
Taylor operators in the integrand of the amplitudes in the BPHZ
formalism \cite{BPHZ}, as far as mathematical rigourousness is
concerned, with the advantage of not modifying the structure of  the
integrand. The divergences are singled out  in terms of internal
momentum integrals which need not be evaluated. Such procedure
generalizes to higher loop order. These so called loop integrals may
be cast solely as a function of an  arbitrary scale which both
parametrises the freedom of separating the divergent part of an
amplitude and plays the role of renormalization group scale.
Accordingly, a minimal  subtraction renormalization scheme in which
such loop integrals are subtracted in the definition of counterterms
naturally emerges. An essential by-product of isolating the
divergencies in IR are surface terms expressed by differences
between loop integrals of the same superficial degree of divergence.
This is a crucial point of IR. Although in principle they are
arbitrary numbers, such surface terms can be shown to be related to
momentum routing invariance in a Feynman diagram, namely the
possibility of a shift in the momentum integration variable.  The
surface terms  play an essential role in preserving symmetries in
Feynman diagram calculations: they translate into finite local
counterterms, whose value is in principle arbitrary but can be
determined  by the symmetry requirements of the underlying theory.

For an account on the connection between gauge invariance and
surface terms in IR see \cite{IR2}-\cite{IR4}; for the study of chiral and gravitational anomalies
and CPT violation see \cite{IR4} and \cite{IR9} in which we also showed that IR is an ideal
arena to display consistently the physics of models which have
finite but undetermined parameters, as discussed by Jackiw in
\cite{JACKIW}. Applications to supersymmetry can be found in \cite{IR7},
where the $\beta$-function of the Wess-Zumino model is calculated to
three loop order and in \cite{IR10}, for the calculation of the
anomalous magnetic moment of the lepton in supergravity.   A
generalization to higher loop order was firstly sketched in
\cite{IR6}. The role of the surface terms, as undetermined parameters
in effective phenomenological models, have been considered in
\cite{IR11}-\cite{IR12}. Along the lines of IR, a complete systematisation of
one-loop four-dimensional Feynman integrals appears in \cite{ORIMAR}.

The purpose of this work is to clarify some issues for calculations
beyond one  loop level within IR  which are basic to any field
theoretical model. In doing so, we pave the  way for  a complete
systematization of IR as an automatic invariant framework order by
order in perturbation theory, i.e. that  preserves gauge invariance
and supersymmetry,  as we have shown at one loop order.

Those issues are: 1) Can we still display the divergencies as basic
loop integrals in  one internal momentum in any loop order,
overlapping divergencies included? 2) In addition,  may such basic
loop integrals be written as  a function of an arbitrary parameter
$\lambda$ which plays the role of renormalization group scale in a
general case  whilst the subtractions are dictated by the BPHZ
forest formula? 3) In connection to this, can the derivatives of the
basic divergent integrals with respect to $\lambda$ also be
displayed by  loop integrals (or constants) in the lines of Implicit
Regularization? 4) How do the surface terms, which encode momentum
routing invariance in the loops, look like in arbitrary loop order?
5) In dimensional methods ultraviolet and infrared divergences
become mixed.  How does IR deal with  this problem, in general, for
one loop and higher loop order amplitudes?

In order to answer these questions we use the simplest
renormalizable scalar  field theory namely $\phi^3$ theory in $6$
dimensions as a working example and show how we  calculate
renormalization group functions in the IR framework. We also outline
the IR rules to work out amplitudes to arbitrary loop order.

\section{Implicit regularization and hidden parameters in basic
divergent integrals at one loop level} \label{IR} \indent In this
section we state the basic steps of IR in one loop calculations. We
also  construct an arbitrary parametrization of the divergences
expressed by loop integrals in order to make contact with other
regularizations. Also we show how the infrared cutoff in the
propagators cancels out in a subtle interplay between the divergent
and finite parts of the amplitude at one loop level, the divergent
part being expressed as a basic divergent integral independent of
the external momenta.  We deal with higher loop in the next section.

\begin{enumerate}

\item  In order to give mathematical rigor to any algebraic 
manipulation performed
in the amplitude, we implicitly assume that a regularization has
been applied. It  can be maintained implicit, the only requirement
being that  the integrand of the amplitude nor the dimension of the
space-time is modified. After eventually performing  Dirac matrix
traces, group index contractions etc.,  we cast the momentum-space
amplitude as a  combination of basic integrals. Typical basic
integrals are: \bq && I, I_\mu, I_{\mu \nu}, \cdots =  \nonumber \\
&& \int \frac {d^{2w}k}{(2\pi)^{2w}} \frac {1, k_\mu, k_\mu k_\nu,
\cdots} {[k+k_1)^2-m^2]\ldots[(k+k_N)^2-m^2]}, \nonumber \eq for a
$N$-point function in a $2w$ integer dimensional space. The internal
momenta $k_i$ are related to the external momenta $p_i$ by $p_N =
k_1 - k_N$, $p_i = k_{i+1} - k_i$, $i = 1, \ldots , N-1$ if we
choose $\sum_{j=1}^{N} p_j = 0$.

\item In the basic integrals, the divergent part is  subtracted  as 
basic divergent integrals which are obtained
by applying recursively the identity
\bq
&&\frac {1}{(p-k)^2-m^2}=\frac{1}{(k^2-m^2)} \nonumber \\
&&-\frac{p^2-2p \cdot
k}{(k^2-m^2)
\left[(p-k)^2-m^2\right]},
\label{ident}
\eq
until the divergent part is free from the external momentum ($p$) 
dependence in the denominator.
This will assure local counterterms. The basic divergent integrals have
the general form
\bq
&& \int^\Lambda \frac{d^{2w}k}{(2 \pi)^{2w}} \frac{g_{\mu_i 
\mu_j}}{(k^2-m^2)^\alpha} , \nonumber \\
&& \int^\Lambda \frac{d^{2w}k}{(2 \pi)^{2w}}
\frac{k_{\mu_1}k_{\mu_2}\cdots k_{\mu_n}}{(k^2-m^2)^\alpha},
\nonumber \eq where  the superscript $\Lambda$ indicates that the
integral is regularized and $w-\alpha\ge $ for the first integral
and $2w-2\alpha+n \ge 0$ for the second. Eventual even powers of
internal momenta in the numerator are simplified  by adding and
subtracting a mass squared term.

\item  Express the basic divergent integrals for which the internal 
momenta carry  Lorentz indices as
a function of surface terms (i.e. integrals of a total divergence). For  
example, in four dimensions:
\bq
&&\int^\Lambda \frac{d^{4}k}{(2 \pi)^{4}} \frac{k_\mu 
k_\nu}{(k^2-m^2)^3} =  \nonumber \\
&& = \frac 14  \int^\Lambda \frac{d^{4}k}{(2 \pi)^{4}} 
\frac{\partial}{\partial k^\nu}
\left( \frac{k_\mu}{(k^2-m^2)^2}\right) + \nonumber \\
&& +   \frac{g_{\mu \nu}}{4}  \int^\Lambda \frac{d^{4}k}{(2
\pi)^{4}} \frac {1}{(k^2-m^2)^2}. \nonumber \eq Such surface terms
are regularization dependent (e.g. they vanish in dimensional
regularization). The possibility of making shifts in the loop
momenta means that the surface terms are null which reveals momentum
routing invariance (MRI). A constrained version of IR (CIR) assumes
that such surface terms are canceled  by local (MRI) restoring
counterterms. In practice, this is automatically realized by setting
them to zero from the start. CIR is an invariant regularization
which preserves the Ward-Slavnov-Taylor identities of the amplitude,
save when quantum symmetry breaking  occurs. In this case they
should be considered as finite arbitrary parameters to be fixed on
physical grounds. That is because anomalies are linked to momentum
routing dependence in Feynman diagram calculations.

\item The basic divergent integrals which encode the ultraviolet 
behavior
of the amplitude need not be evaluated. We adopt the following
notation: \be I_{log}(m^2)=  \int^\Lambda \frac{d^{2w}k}{(2
\pi)^{2w}}\frac{1}{(k^2-m^2)^w} \label{ilog} \ee \be I_{quad}(m^2)=
\int^\Lambda \frac{d^{2w}k}{(2 \pi)^{2w}} \frac{1}{(k^2-m^2)^{w-1}},
etc. \label{iquad} \ee for logarithmically and  quadratically basic
divergent integrals, etc. .

\item These objects  can be subtracted as they stand in the definition 
of renormalization functions through, for instance,
the definition of local counterterms in the process of
renormalization. A minimal, mass independent scheme is defined by
substituting $m^2$ with $\lambda^2 \ne 0$  using the regularization
independent relation in $2w$ dimensional space-time.
\be
\label{scaleg} I_{log}(m^2)=I_{log}(\lambda^2)+b_{2w}
(-1)^{w}\ln{\left( \frac{\lambda^2}{m^2}\right)},
\ee
where $b_{2w}=\frac{i}{(4\pi)^{w}}\frac{1}{\Gamma(w)}$, and subtracting
$I_{log}(\lambda^2)$, $\lambda$ playing the role of renormalization
group scale in the renormalization group equation. For infrared safe
massless models a systematic cancelation of $ln(m^2)$ steming from
(\ref{scaleg}) and from ultraviolet finite part will render the
amplitude well defined as $m^2 \rightarrow 0$.

\item The remaining ultraviolet finite integrals are evaluated as
usual, using Feynman parameters or an extensive library of methods
in the momentum space \cite{SMIRNOV}.

\end{enumerate}

A systematic presentation of the finite part of  one loop $N$-point 
Green's functions in four dimensions is given in \cite{ORIMAR}.
\vskip1.0cm
As a matter of illustration consider the self energy graph of the  
massless  $\phi^3_6$  theory as an example,
which is power counting quadratically divergent,
\be
-i \Sigma(p^2)=\frac{g^2}{2}\int^\Lambda_k
\frac{1}{(k^2-m^2)
\left[(p-k)^2-m^2\right]}.
\ee
From now on we adopt the notation $\int_k^\Lambda \equiv  \int^\Lambda 
d^{2w}k/(2 \pi)^{2w}$ in  $2w$ dimensions. An infrared cut off is 
introduced in the propagators and the limit $m^2 \to 0$ will be taken at the 
end of the
calculation. In order to separate the divergences, the identity
(\ref{ident}) is applied three times to yield
\bq
&& -i\Sigma(p^2) = \frac{g^2}{2}\int_k^\Lambda \Bigg\{ 
\frac{1}{(k^2-m^2)^2}
-\frac{p^2}{(k^2-m^2)^3} \nonumber \\&& + \frac{4(p \cdot
k)^2}{(k^2-m^2)^4} + \frac{p^4}{(k^2-m^2)^4} \nonumber \\
 && - \frac{(p^2-2p \cdot
k)^3}{(k^2-m^2)^4
\left[(p-k)^2-m^2\right]}\Bigg\},
\label{SE}
\eq

Note that the first three integrals on the r.h.s. are divergent and 
regularization dependent.
The first one is quadratically whereas the other two logarithmically 
divergent integrals.
The remaining integrals are ultraviolet (UV) finite  and can be 
evaluated using Feynman parameters.
However, in a massless case they are still  infrared divergent.  In the 
sense of IR, in which
we do not evaluate loop integrals, we will show that a cancellation of 
the infrared divergences
coming from the UV divergent and finite parts will always take place 
before taking the limit $m^2 \to 0$.

Let us start by discussing the regularization dependent terms. The 
basic one-loop
regularization dependent objects of a six
dimensional  theory are obtained by choosing $w=3$ in eq. (\ref{ilog}) 
and (\ref{iquad}).

In equation (\ref{SE}) there appears, besides the quadratic divergence 
which vanishes
for a massless theory if a adequate parametrization is used, a 
difference between logarithmic
divergent integrals. Let us write $-i \Sigma = -i\Sigma_\infty -i 
\Sigma_F$, ($-i\Sigma_\infty$) $-i\Sigma_F$
standing for the power counting (divergent) finite basic integrals. 
Then according to the rules of IR we have
\bq
\frac{-i\Sigma_\infty}{g^2} &=& 2 p^\mu p^\nu \int_k^\Lambda 
\frac{k_\mu
k_\nu}{(k^2-m^2)^4}-\frac{p^2}{2}I_{log}(m^2) \nonumber \\
&=& 2 p^\mu p^\nu (\Upsilon^0_{\mu \nu} - \frac{1}{12} g_{\mu \nu} 
I_{log}(m^2))\, ,
\label{DIR}
\eq
with
\bq
 \Upsilon^0_{\mu \nu} &=& \int_k^\Lambda \frac{k_\mu 
k_\nu}{(k^2-m^2)^4} - \frac
{g_{\mu \nu}}{6}I_{log}(m^2) \nonumber \\
&=& -\frac 16 \int_k^\Lambda \frac{\partial}{\partial k^\mu}
\left( \frac{k_\nu}{(k^2-m^2)^3}\right).
\label{ST1}
\eq
whereas $\Sigma_F$ can be easily evaluated and has a simple form in the 
limit $m^2 \downarrow 0$. If we write
\be
\Upsilon_{\mu \nu}=a g_{\mu \nu}
\ee
$a$ being an arbitrary constant, (\ref{SE}) reads
\bq
\frac{-i \Sigma (p^2)}{g^2} &=& 2 a p^2 - \frac{p^2}{6} I_{log} (m^2) + 
\nonumber \\
&-& \frac{b}{6} p^2 \ln \Big( -\frac{p^2}{m^2}\Big)  + \frac{4 b}{9} 
p^2 \, ,
\eq
with $b=i/[2(4\pi)^3]$. Using relation (\ref{scaleg}) for six 
dimensions,
$I_{log}(m^2)=I_{log}(\lambda^2)-b\ln{\left(\frac{\lambda^2}{m^2}\right)}$, 
$\lambda^2\ne0$.
In the equation above, it is clear that the divergent logarithms as 
$m^2\downarrow 0$ cancel out
as we indicated before. We show in the next sections that such infrared 
divergence cancellation
mechanism takes place for arbitrary $N$-point functions at one loop 
order as well as for higher
loop order (see also \cite{IR7}). The final answer is
\bq
-i\Sigma(p^2)&=&-\frac {p^2g^2}{6} \Bigg( I_{log}(\lambda^2)+b
 \ln{\left(-\frac{p^2}{\lambda^2}\right)} \nonumber \\
 &-& \frac 83 b \Bigg),
 \label{eq:se1}
\eq
in which we have taken $a=0$. It is evident that the momentum routing 
invariance requirement $a=0$
is not essential for such simple scalar theory. In fact the choice of 
the routing in this case contributes
as a finite local counterterm which is fixed in the definition of the 
renormalization scheme, or equivalently
in our case in a redefinition of the arbitrary constant $\lambda$. 
Within IR, the ultraviolet behaviour is
displayed in terms of basic divergent integrals such as 
$I_{log}(\lambda^2)$ above, etc. . Their generalization
to higher loop order as their (regularization dependent) explicit 
evaluation may contaminate the underlying
physics by breaking symmetries and introducing junk constants. The 
derivatives of the basic divergent integrals
with respect to the scale $\lambda$ (which are important for the 
calculation of renormalization group functions)
are either finite or may be written in the form of  basic divergent 
integrals being therefore regularization
independent as well. The simple subtraction of $I_{log}(\lambda^2)$ in 
the definition of renormalization constants
defines a minimal mass independent scheme in IR. In section 
\ref{secnloop} we calculate the renormalization group functions.

Before proceeding to $N$-point functions and higher loop order, let us 
make an interlude in order to illustrate the
role of the surface terms, which encode a  specific momentum routing in 
the loops as we discussed in the introduction,
on gauge symmetry.  The generalization to higher loop order is subject 
of a forthcoming contribution.

The one-loop QED vacuum polarization tensor in $4D$, as it is 
calculated in
\cite{IR3} with arbitrary momentum in the internal lines, $k_1$ and 
$k_2$:
\bq
\Pi_{\mu \nu}&=&\Pi(p^2)(p_\mu p_\nu-p^2 g_{\mu \nu}) \nonumber \\
&+& 4\left( \alpha_1 g_{\mu\nu}-\frac{1}{2}(k_1^2+k_2^2)\alpha_2 g_{\mu
\nu} \right. \nonumber \\
&+& \left. 
\frac{1}{3}(k_{1}^{\alpha}k_{1}^{\beta}+k_{2}^{\alpha}k_{2}^{\beta}
+k_{1}^{\alpha}k_{2}^{\beta})
\alpha_3 g_{\{\mu \nu}g_{\alpha \beta\}}  \right.\nonumber \\ &-& 
\left.
(k_1+k_2)^{\alpha}(k_1+k_2)_{\mu}\alpha_2 g_{\nu
\alpha} \right. \nonumber \\
&-& \left. 
\frac{1}{2}(k_1^{\alpha}k_1^{\beta}+k_2^{\alpha}k_2^{\beta})g_{\mu \nu}
\alpha_2 g_{\alpha \beta}  \right).
\label{QED}
\eq
In the equation above, $p=k_1-k_2$ is the external momentum and
$$
\Pi(p^2)= \frac{4}{3} \Big[ I_{log}(\lambda^2) + \frac{i}{(4 \pi)^2} 
\ln\Big( \frac{e^2 \lambda^2}{-p^2}\Big)- \frac{i}{3 (4 \pi)^2}\Big]
$$
includes the basic divergent integral. We have  chosen the massless 
limit just for the sake of simplicity. Now the momentum routing dependent 
terms are proportional to $\alpha_i$'s, namely
\be
\alpha_1 g_{\mu \nu} \equiv \int^{\Lambda}_k
\frac{g_{\mu\nu}}{k^2-m^2}-
2\int^{\Lambda}_k
\frac{k_{\mu}k_{\nu}}{(k^2-m^2)^2},
\ee
\be
\alpha_2 g_{\mu \nu} \equiv \int^{\Lambda}_k
\frac{g_{\mu\nu}}{(k^2-m^2)^2}-
4\int^{\Lambda}_k
\frac{k_{\mu}k_{\nu}}{(k^2-m^2)^3}
\label{CR1}
\ee
and
\bq
\alpha_3 g_{\{\mu \nu}g_{\alpha \beta\}}  & \equiv &
g_{\{\mu \nu}g_{\alpha \beta \}}
\int^{\Lambda}_k
\frac{1}{(k^2-m^2)^2} \nonumber \\
&-&24\int^{\Lambda}_k
\frac{k_{\mu}k_{\nu}k_{\alpha}k_{\beta}}{(k^2-m^2)^4}.
\label{CR2}
\eq
These parameters are surface terms. It can be easily shown that
\be
\alpha_2 g_{\mu \nu}= \int_k ^\Lambda \frac{\partial}{\partial k^\mu}
\left( \frac{k_ \nu}{(k^2-m^2)^2} \right),
\ee
\be
\alpha_1 g_{\mu \nu}= \int_k ^\Lambda \frac{\partial}{\partial k^\mu}
\left( \frac{k_ \nu}{(k^2-m^2)} \right)
\ee
and
\be
\int_k^\Lambda \frac{\partial}{\partial k^\beta}
\left[ \frac{4k_\mu k_\nu k_\alpha}{(k^2-m^2)^3} \right]
=g_{\{\mu \nu}g_{\alpha \beta\}}(\alpha_3-\alpha_2).
\ee
In the case of gauge symmetry, both abelian and nonabelian, a 
constrained version of IR in which such surface terms are set to vanish delivers 
gauge invariant amplitudes automatically \cite{IR8}, which is 
illustrated in the simple example above . The cancellation of these surface 
terms is also a requirement to preserve Supersymmetry \cite{IR10}. That 
is to say: they represent the symmetry restoring local counterterms.  In 
other words, momentum routing invariance seems to be the crucial 
property in a Feynman diagram in order to preserve symmetries. In fact such 
surface terms evaluate to zero should we employ DREG to explicitly 
evaluate them. This property somewhat reveals why DREG is manisfestly gauge 
invariant yet it breaks supersymmetry because invariance of the action 
with respect to supersymmetry transformations only holds in general for 
specific values of
the space-time dimension. \footnote{ The idea of associating momentum 
routing in the loops with symmetry properties of the Green's functions 
has been exploited in a framework named Preregularization which did not 
call for momentum routing invariance but instead fixed the routing in 
order to fulfill certain Ward identities. \cite{MCKEON}}.

A particular situation, however, is the ocurrence of quantum symmetry 
breakings (anomalies) . Anomalies within perturbation theory may present 
some oddities such as preserving a certain symmetry at the expense of 
adopting a special momentum routing in a Feynman diagram e.g. in the 
(Adler-Bardeen-Bell-Jackiw) AVV triangle anomaly \cite{CURRENT}.  In the case 
of chiral anomalies, IR has been shown to preserve the democracy 
between the vector and axial sectors of the Ward identities which is a good 
'acid test' for regularizations. The arbitrary parameter 
represented by the surface term remains undetermined and floats between the 
axial and vector sectors of the Ward identities. That is to say, in the 
anomalous amplitudes, there is
no possibility of restoring, at the same time, the axial and the 
vectorial Ward
identities. The counterterm that will restore one symmetry causes the 
violation of the other and therefore  it does not make sense to set the 
surface terms to zero. The answer is to be established by physical 
constraints on such amplitude. This feature has also been illustrated  in 
the description of two-dimensional gravitational anomalies \cite{IR9}.

We end this section  by  showing that  we can parametrize
the basic divergent integrals should we wish to make contact with an
explicit regularization without assigning a definite value to a
regularization dependent parameter hidden in an UV divergent amplitude.

As we have argued before, in IR we need only the
(loop) integral representation of the divergence and its derivatives. 
The latter can always be
cast as a loop integral as well. Alternatively, we could construct a 
general parametrization
for such basic divergent integrals, which we do here for the purpose of 
illustration and comparison
with other regularization methods. It is however not necessary for 
calculational purposes.

For instance the basic logarithmically divergent integral at one loop 
order satisfies
\be
\frac {\partial}{\partial m^2}I_{log}(m^2)=\frac{b}{m^2},
\ee
with $b=i/[2(4\pi)^3]$, from which  we may construct a general 
parametrisation
by integrating the equation above. Namely
\be
I_{log}(m^2)=-b \ln{\left( \frac{\Lambda^2}{m^2}\right)}+\beta,
\label{pilog}
\ee
where $\beta$ is an arbitrary constant and $\Lambda$ is a mass
parameter introduced for dimensional reasons, which dictates the UV 
behavior of the
integral. It is very important to notice that a free arbitrary 
parameter $\beta$ appears.
The explicit parametrization of the differences between logarithmically 
divergent integrals
(\ref{CR1}) and (\ref{CR2}) which represents the surface terms will 
render an arbitrary constant
in the end as the ultraviolet divergent logarithms are cancelled.

Now, because
\be
\frac {\partial}{\partial m^2}I_{quad}(m^2)=2I_{log}(m^2),
\ee
a possible parametrization of $I_{quad}(m^2)$ is
\be
I_{quad}(m^2)= -\frac{i}{(4\pi)^3} m^2  \left[ \ln{\left(
\frac{\Lambda^2}{m^2}\right)}+\beta' \right].
\label{piquad}
\ee
Let us now evaluate the quadratic divergence in (\ref{iquad}). 
Recalling
that In DREG, for dimensional reasons, we have a factor
$\mu^{2\epsilon}$, where $\mu$ is a mass parameter,
$\epsilon=6-d$, $d$ is  the space-time dimension, we have
\bq
\mu^{2 \epsilon}I_{quad}(m^2)&=& -\frac {i}{(4\pi)^3}m^2\Bigg( \frac 
{1}{\epsilon} + 1 +
\ln{\left(\frac{-4\pi \mu^2}{m^2}\right)}\nonumber \\ &+&\gamma_E+ 
{\cal
O}(\epsilon)\Bigg),
\eq
which vanishes as  $m^2 \to 0$ and resembles the parametrization  
(\ref{piquad}).
Now, in order to compare IR with DREG let us calculate (\ref{DIR}). 
Because
\be
\mu^{2\epsilon}I_{log}(m^2)=-b\left[ \frac {1}{\epsilon} +
\ln{\left(-\frac{4\pi \mu^2}{m^2}\right)}+ \gamma_E +{\cal
O}(\epsilon)\right] \, ,
\ee
and  that $\Upsilon^{0 \,\, DREG}_{\mu \nu} =0$, we obtain
  \bq
&& \mu^{2\epsilon}\frac{\Sigma_{\infty}}{g^2} = -\mu^{2\epsilon}\frac 
16 p^2 I_{log}(m^2) =
\nonumber \\&&
-\frac{b}{6} \Bigg[ \frac {1}{\epsilon} +
\ln{\Bigg(-\frac{4\pi \mu^2}{m^2}\Bigg)}+ \gamma_E +{\cal
O}(\epsilon)\Bigg],
\eq
which should be compared with (\ref{DIR}).

\section{One loop n-point functions}
\indent

We now turn ourselves to a general one-loop amplitude for a general 
massless
amplitude, using the procedure described above.

In a theory defined in $d$ dimensions, with $d$ even, one will
frequently deal with logarithmically ultraviolet divergent integrals of
the
type,
\bq
&& I=\int^\Lambda \frac {d^{2w}k}{(2\pi)^{2w}}
\frac{1}{(k^2-m^2)[(k+p_1)^2-m^2]}\cdots \nonumber \\ && \cdots
\frac{1}{[(k+p_n)^2-m^2]},
\eq
where $n=w-1=d/2-1$. The following expansion can
be performed,
\be
I=\int_k^\Lambda
\frac{1}{(k^2-m^2)^n[(k+p_n)^2-m^2]}-\sum_{i=1}^{n-1}I_i,
\ee
with
\bq
&& I_i=\int_k \frac{p_i^2+2p_i\cdot
k}{(k^2-m^2)^{i+1}[(k+p_i)^2-m^2]} \nonumber \\ && \cdots
\frac{1}{[(k+p_n)^2-m^2]}.
\eq
The expansion above has been obtained  by using the
identity (\ref{ident}) in all factors containing external momenta
$p_i$ dependence from $i=1$ to  $i=n-1$.  Each time the identity is
applied, we obtain a
new integral $I_i$ that is ultraviolet finite. This is why we do not
use the index $\Lambda$ in
these integrals.
Next we want to show that in the sense of IR the limit $m^2 \to 0$ is 
well defined.
Our proof is done in two parts: first we show
that the integrals $I_i$ are well defined . A few algebraic 
manipulations which will lead
to eq. (\ref{eq32}) will make that clear. Then we show how to handle 
the $m^2$ dependence in the regularized integral.

We begin the calculation of $I_i$ by using  Feynman's parametrization
\bq
&& \frac{1}{a_1\cdots a_l^\alpha}=\frac{(l+\alpha-2)!}{(\alpha-1)!}
\int d X \nonumber \\ &&
\frac{(1-x_1-\cdots
-x_{l-1})^{\alpha-1}}{[(a_1-a_l)x_1+(a_2-a_l)x_2+
\cdots +a_l]^{l+\alpha-1}} \nonumber,
\eq
where
\be
\int dX \equiv \int _0^1 dx_1\int _0^{1-x_1}dx_2\cdots \int
_0^{1-x_1-\cdots -x_l}dx_{l-1}.
\ee
In the integral $I_i$ we have $\alpha=i+1$ and $l=w-i+1$, so that
\bq
&& I_i=\frac {w!}{i!}\int dX(1-x_1-\cdots -x_{w-i})^i \times \nonumber
\\
&& (p_i^2-2p_i^2  x_{w-i}-2 (p_i\cdot p_{i+1}) x_{w-i-1}\cdots-2 
(p_i\cdot
p_n) x_1) \nonumber \\ &&
\times \int_k \frac {1}{(k^2+Q^2)^{w+1}},
\eq
with
\bq
&& Q^2=p_n^2x_1(1-x_1)+\cdots +p_i^2x_{n-i+1}(1-x_{n-i+1}) \nonumber \\
&& -
2\sum_{l\neq t}(p_l \cdot p_t) x_{n-l+1}x_{n-t+1} -m^2.
\eq
The final result is given by
\bq
&& I_l=\frac {i}{(4\pi)^w} \frac {(-1)^w}{l!}
\int dX(1-x_1-\cdots -x_{w-l})^l \nonumber \\
&& (p_l^2-2 p_l^2 x_{w-l}-2 (p_l\cdot p_{l+1}) x_{w-l-1}\cdots-2 
(p_l\cdot
p_n) x_1) \nonumber \\ && \times \frac{1}{Q^2}.
\label{eq32}
\eq
It is clear from the expression above that whenever the external
momenta
are such that $p_i^2\neq 0$, the
limit $m^2 \to 0$ is well defined.

We are now left with the regularization dependent integral,
\be
I^\Lambda=\int_k^\Lambda \frac{1}{(k^2-m^2)^n[(k+p_n)^2-m^2]},
\ee
from which we can separate the appropriate basic divergence
$I^d_{log}(m^2)$ (typical of this
dimension) from its finite part:
\be
I^\Lambda=I^d_{log}(m^2)-
\int_k \frac {p_n^2-2p_n\cdot k}{(k^2-m^2)^w[(k+p_n)^2-m^2]},
\ee
with $I^d_{log}(m^2)$ given by eq. (\ref{ilog}),
and where we must remember that $n=w-1$. Note that the cutoff mass
$m^2$ is present in both finite and divergent parts. We will now
show that, due to a scale relation, the $m^2$ dependence of the
divergent integral can be extracted and precisely cancels out the
$m^2$ dependence of the finite part.
First, we make use of the identity
\bq
&& \frac {1}{(k^2-m^2)^w}=\frac {1}{(k^2-\lambda^2)^w}
-(\lambda^2-m^2) \nonumber \\
&&\times \sum_{i=1}^w \frac
{1}{(k^2-m^2)^i(k^2-\lambda^2)^{w-i+1}},
\eq
and obtain the $d$- dimensional scale relation (\ref{scaleg}),
\be
I^d_{log}(m^2)=I^d_{log}(\lambda^2)+b_d (-1)^{d/2}\ln{\left(
\frac{\lambda^2}{m^2}\right)},
\ee
with $b_d=\frac{i}{(4\pi)^{d/2}}\frac{1}{\Gamma(d/2)}$. Again, like in
the six dimensional case, the relation above could be obtained from a
particular
regularization ,
but, as we have seen, the relation is regularization independent.
On the other hand, the calculation of the finite part
yields:
\bq
&& \int_k \frac {p_n^2-2p_n\cdot k}{(k^2-m^2)^w[(k+p_n)^2-m^2]} 
\nonumber \\
&&=(-1)^{d/2}
b_d\left(\frac d2-1\right)  \int_0^1dx(1-x)^{d/2-2} \nonumber \\
&&\ln {\left( \frac{p_n^2x(1-x)-m^2}{(-m^2)}\right)}.
\eq
We can now clearly see, by collecting the two parts together, that
\bq
&& I^\Lambda= I^d_{log}(\lambda^2)-(-1)^{d/2}b_d\left(\frac
d2-1\right)\int_0^1 d x \nonumber \\ && (1-x)^{d/2-2}
\ln {\left( \frac{p_n^2x(1-x)}{(-\lambda^2)}\right)},
\eq
where the limit $m^2\to 0$ has been taken. The whole integral,
\be
I=I^\Lambda-\sum _{i=1}^{n-1}I_i,
\label{1l}
\ee
becomes then
$m^2$ independent. This result, in conjunction with equation 
(\ref{eq32}), will be
important for the generalization of IR to higher loop order in order to 
subtract subdivergences. Also, once
the subtraction scheme is appropriately chosen, the object
$I^d_{log}(\lambda^2)$
can be subtracted without having to be explicitly evaluated.
The calculation above establishes a procedure for
dealing with massless theories at one loop order in the context of
IR. This procedure can be straightforwardly adapted to higher orders
when non-overlapping divergencies occur, as shown  in the ref.
\cite{IR7},
where the $\beta$-function of the massless Wess-Zumino model was
calculated at three loop order. In that contribution, it was shown that
a $n$-loop scale relation can be used in order to appropriately handle
the dependence on the infrared cut-off.

We shall now generalize this procedure
 for the general case
when overlapping divergencies occurs with support in the
BPHZ-forest formula.

\section{Higher order calculations}
\label{secnloop}
\indent

Renormalization is a recursive program. In a $n$-loop order calculation 
the main goal is to identify the typical divergence of the 
$n^{th}$order  and the finite part of an amplitude once the $(n-1)^{th}$ has been 
renormalized.
In \cite{IR7}, we have calculated within IR the three-loop $\beta$
function for the Wess-Zumino model in which there were not overlapping 
divergences to consider.

In fact IR can be applied in the sense of defining
loop integrals as basic ultraviolet divergent objects also when 
overlapping divergences occur. A regularization independent scale relation 
will also emerge in this case which serves to both cancel infrared 
divergences in a similar fashion as we presented in the last sections and 
introduce a renormalization group scale. New surface terms appear beyond 
one loop level and are expected to play an essential role in preserving 
gauge symmetry. Moreover IR rules are  compatible with BPHZ forest 
formula,  which judiciously defines the set of subtractions to 
remove the subdivergences. The rules to implement IR beyond one loop 
order can be simply stated once we adopt the version of  the forest 
formula which is analogous to the ordinary counterterm method in which the 
subtraction operators are translated into a local counterterm which 
substitutes the subgraph.  This is implemented via a recursion equation 
which involves disjoint renormalization parts only \cite{BPHZ}, described in 
texbooks, for instance \cite{MUTA}.

\begin{enumerate}

\item Starting from one loop order, the (sub)divergences should be 
expressed in terms of basic divergent integrals, which can be written
in terms of one internal momentum only and an arbitrary non-vanishing 
parameter $\lambda$, the renormalization
group scale of the method. For this purpose identity  (\ref{ident}) is 
judiciously used.  We assume that
subdiagrams are proper (one particle irreducible) which is sufficient 
to discuss renormalization in general.

\item The counterterms are defined order by order through a minimal 
subtraction process within implicit regularization
which amounts to subtracting the (sub)divergences as basic divergent 
integrals as defined  above

\item The subtraction of subdivergences follows the forest formula 
written in a way which is equivalent
of the subtractions of local counterterms only. That is to say given a 
diagram $G$  the subtraction operator
for a certain renormalization part  $H$ has the effect of crushing $H$ 
to a point and multiplying the counterterm
to the resulting  Feynman integral.  In  this recursive process only 
the set of disjoint renormalization parts is needed.

\end{enumerate}

Mathematically speaking, a recursion relation for the subtraction of
subdivergencies based on the forest formula for a Feynman diagram $G$ 
in which only the disjoint renormalization parts are needed can be 
written as
\bq
\bar{R}_G F_G &=& \sum_\psi F_{G/\psi} \prod_{H \in \psi} 
(-t^H\bar{R}_H F_H)) \nonumber \\
&=& \sum_{\psi} \prod_{H \in \psi} (-t^H \bar{R}_H) F_G
\eq
where $\bar{R}_G$ is the operation to subtract only subdivergencies, 
$F_{G(H)}$ is the part of the amplitude which represents the (sub)graph 
$G(H)$, $\psi$ is a set of disjoint renormalization parts of $G$, namely
$$
\psi = \{ H/H \subset G, H = \mbox{proper, disjoint}, d_H \geq 0 \} \, 
,
$$
 $G/\psi$ representing the diagram obtained from $G$ by crushing all 
$H$ in $\psi$ to a point. The counterterm graph so obtained can be 
constructed with the loop integrals characteristic of IR order by order. Of 
course a further operation $1-t_G$ is required to the overall 
divergence. The natural (minimal)  renormalization scheme in IR is to subtract 
the basic divergent loop integrals as a function of the arbitrary scale 
$\lambda$.

\begin{figure}[h,t]
\centerline{\hbox{
  \epsfxsize=1.8in
  \epsfysize=0.9in
  \epsffile{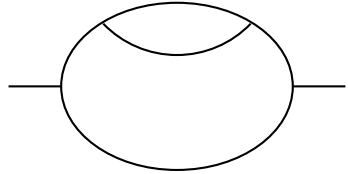}
    }
 }
  \caption{The 2-loop diagram with a nested subdivergence for the $\phi
^3_6$ self-energy
         }
  \label{fig1}
\end{figure}

For the massless case the basic divergent integrals are very simple. 
For instance the logarithmic basic divergence to $n$-loop order is given 
by
\bq
\label{Ilogn}
&& I^{(n)d}_{log}(m^2)=\int^\Lambda \frac{d^{2w}k}{(2\pi)^{2w}}
\frac{1}{(k^2-m^2)^w} \times \nonumber \\ && \ln
^{n-1}{\left(-\frac{(k^2-m^2)}{\lambda^2}\right)}
\eq
and the corresponding scale relation is written as
\be
I^{(n)d}_{log}(m^2)=I^{(n)d}_{log}(\lambda^2)+ \sum_{i=0}^n a^{(w)}_i
\ln^i{\left(\frac{m^2}{\lambda^2}
\right)},
\ee
with the $a^{(w)}_i$ depending on the dimension $2w$.
\vskip1.0cm
Let us consider the two-loop contribution to the self-energy
in $\phi^3_6$ theory.
The nested subdiagram in fig. (\ref{fig1}) can be calculated by simply 
substituting the subdiagram by its finite part. The remaining integral, 
in one internal momentum, includes only the divergence of the order and 
the finite part.

The amplitude for the diagram depicted in fig. (\ref{fig1}) reads
\be
-i\Sigma_1^{(2)}(p^2)=i \frac{g^4}{2}\int_k \frac{1}{k^4(k-p)^2}
\int_l \frac{1}{l^2(l-k)^2}.
\ee
The integral in the momentum $l$ has been calculated before (eq. 
(\ref{eq:se1})):
\be
I=-\frac{k^2}{3}\left\{ I_{log}(\lambda^2)+b\ln { 
\left(-\frac{k^2}{\lambda^2}\right)}
-\frac 83 b\right\}.
\label{eq:I}
\ee
We take its finite part and obtain:
\bq
&&-i\bar \Sigma_1^{(2)}(p^2)= -i \frac{g^4b}{6}\int_k 
\frac{1}{k^2(k-p)^2}\times \nonumber \\
&&\left( \ln { \left(-\frac{k^2}{\lambda^2}\right)}
-\frac 83 \right) \nonumber \\
&&= -i \frac{g^4b}{6}\left(I^{(2)}- \frac 83 I\right),
\label{eq:II}
\eq
where
\be
I^{(2)}=\int_k \frac{1}{k^2(k-p)^2}
\ln { \left(-\frac{k^2}{\lambda^2}\right)} \, .
\ee
Notice that
\be
-i\bar \Sigma_1^{(2)}(p^2) = (1-t)(-i \Sigma_1^{(2)}(p^2))\, ,
\ee
where $-t$ corresponds to the subtraction operator which has the effect 
of crushing the nested subdivergence
to a point and multiplying the counterterm, which has been calculated 
in the previous order, to the resulting Feynman integral.

\begin{figure}[h,t]
\centerline{\hbox{
 \epsfxsize=2in
   \epsfysize=0.9in
   \epsffile{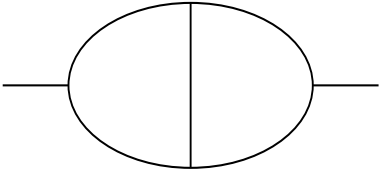}
     }
  }
  \caption{The 2-loop overlapping self-energy
         }
  \label{fig2}
\end{figure}

\begin{figure}[h,t]
\centerline{\hbox{
   \epsfxsize=3in
   \epsfysize=1in
   \epsffile{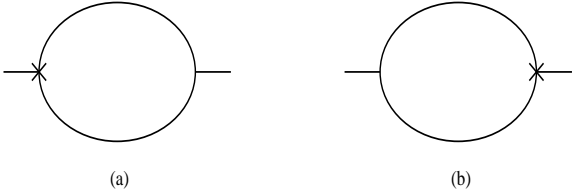}
     }
  }
  \caption{Counterterms to cancel the subdivergences of the overlapping
diagram
          }
  \label{fig3}
\end{figure}

Now we evaluate $I^{(2)}$.  We introduce an infrared cutoff $m^2$ as we 
have done before to rewrite
\be
I^{(2)}=\int_k \frac{1}{(k^2-m^2)[(k-p)^2-m^2]}
\ln { \left(\frac{k^2-m^2}{-\lambda^2}\right)} \, ,
\ee
which is ultraviolet quadratically divergent. The identity expressed by 
eq. (\ref{ident})
is applied three times to yield
\bq
&&I^{(2)}=\int_k^\Lambda  \ln { 
\left(\frac{k^2-m^2}{-\lambda^2}\right)}
 \Bigg\{ \frac{1}{(k^2-m^2)^2} \nonumber \\
&&-\frac{p^2}{(k^2-m^2)^3} + \frac{4(p \cdot
k)^2}{(k^2-m^2)^4} + \frac{p^4}{(k^2-m^2)^4} \nonumber \\
 && - \frac{(p^2-2p \cdot
k)^3}{(k^2-m^2)^4
\left[(p-k)^2-m^2\right]}\Bigg\}.
\eq
We define the basic
two-loop divergent integrals in $6$ dimensions,
\be
I_{log}^{(2)}(m^2)= \int^\Lambda_l \frac {1}{(l^2-m^2)^3}
\ln{\left(-\frac {(l^2-m^2)}{\lambda^2}\right)}
\ee
and
\be
I_{quad}^{(2)}(m^2)=\int^\Lambda_l \frac {1}{(l^2-m^2)^2}
\ln{\left(-\frac {(l^2-m^2)}{\lambda^2}\right)} .
\ee
Clearly  $I_{quad}^{(2)}(m^2)$   vanishes for massless theories in a 
regularization independent fashion.
whilst the divergent basic integrals with  Lorentz indices can be 
written in function of a surface term:
\bq
&&\Theta_{\alpha \beta}(m^2)=\int^\Lambda_k \frac {k_\alpha
k_\beta}{(k^2-m^2)^4}
\ln{\left(-\frac {(k^2-m^2)}{\lambda^2}\right)} \nonumber \\
&&= \frac{1}{6} \left\{ g_{\alpha \beta}\left( I_{log}^{(2)}(m^2)+ 
\frac 13 I_{log}(m^2)\right)
\right. \nonumber \\
&&\left. - \int_k \frac{\partial} {\partial k^\beta}
\left( \frac {k_\alpha}{(k^2-m^2)^3}\ln{\left(-\frac 
{(k^2-m^2)}{\lambda^2}\right)}\right)
\right\}.\nonumber \\
&&
\eq
The surface term will be set to zero for the sake of momentum routing 
invariance.  Thus we have
\be
I^{(2)}=- \frac {p^2}{3}I_{log}^{(2)}(m^2) +\frac 29 p^2 I_{log}(m^2)
+ \tilde I^{(2)},
\label{eq:I(2)}
\ee
$\tilde I^{(2)}$ being the finite part which is given by
\bq
&& {\tilde{I}}^{(2)} =  \int_k  \frac{p^4}{(k^2-m^2)^4} \ln { 
\left(\frac{k^2-m^2}{-\lambda^2}\right)} - \nonumber \\
&& \int_k \frac{(p^2-2p \cdot
k)^3}{(k^2-m^2)^4
\left[(p-k)^2-m^2\right]}\ln { 
\left(\frac{k^2-m^2}{-\lambda^2}\right)}.
\eq
The finite integrals above can be easily evaluated using the identity,
\be
\ln{a} = \lim_{\epsilon \to 0} \frac {1}{\epsilon}
\left(a^\epsilon-1\right),
\label{ident2}
\ee
which after Feynman parametrization in the limit where $m^2 \to 0$ is 
given by
\bq
&&\tilde I^{(2)} = \frac{b}{18} p^2 \left\{ \ln{\left(\frac 
{m^2}{\lambda^2}\right)}
\left[ 6 \ln{\left(-\frac {p^2}{m^2}\right)} -16\right] \right. 
\nonumber \\
&& + \left. 3 \ln ^2{\left(-\frac {p^2}{m^2}\right)}
-11 \ln{\left(-\frac {p^2}{m^2}\right)} +11 \right\}.
\eq
We see the infrared divergence, parametrized by $m^2$, appearing in the 
divergent
and in the finite part. In order to observe the cancellation of these 
two contributions,
we use the one-loop (\ref{scale}) and the two-loop scale relation,
\bq
\label{scale2l}
&& I_{log}^{(2)}(m^2)=I_{log}^{(2)}(\lambda^2)+ b\left\{ \frac 12 
\ln^2{
\left(\frac{m^2}
{\lambda^2}\right)} \right. \nonumber \\
&& \left. + \frac 32
\ln{\left(\frac{m^2}{\lambda^2}\right)}\right\},
\eq
so that eq. (\ref{eq:I(2)}) becomes
\bq
&&I^{(2)}=\frac {p^2}{3}\left\{ - I_{log}^{(2)}(\lambda^2)
+\frac 23 p^2 I_{log}(\lambda^2) \right. \nonumber \\
&& \left. -\frac b2 \ln ^2{\left(-\frac {p^2}{\lambda^2}\right)}
+ \frac{11}{6} b \ln{\left(-\frac {p^2}{\lambda^2}\right)} 
+\frac{11}{6} b \right\}.
\label{eq:III}
\eq
Now equations (\ref{eq:I}), (\ref{eq:II}) and (\ref{eq:III}) together 
give
\bq
&&-i \bar \Sigma_1^{(2)}(p ^2)=\frac{ig^4 b p^2}{18}
\left\{  I_{log}^{(2)}(\lambda^2)
-\frac {10}{3} I_{log}(\lambda^2) \right. \nonumber \\
&& \left. +\frac b2 \ln ^2{\left(-\frac {p^2}{\lambda^2}\right)}
- \frac{27}{6} b \ln{\left(-\frac {p^2}{\lambda^2}\right)} 
+\frac{95}{18} b \right\}.
\eq
\vskip1.0cm
Let us now consider the
two-loop overlapping contribution for the self-energy in $\phi_6^3$ 
theory, represented
by fig. (\ref{fig2}).
It is given by
\bq
&&-i \Sigma_2^{(2)}(p^2)= i\frac{g^4}{2}A \nonumber \\
&&= i\frac{g^4}{2} \int_{k,l}^{\Lambda}
\frac{1}{k^2 l^2 (k-l)^2 (p-k)^2 (p-l)^2}.
\eq
To solve the overlapping integral $A$, it will be easier to rewrite it 
as
$$A=-A_1+A_2+{\mbox{surface\,\, term}},$$
by using the identity \cite{FIELDS}:
\bq
&&\frac {\partial}{\partial k_\alpha}\left( 
\frac{(k-l)_\alpha}{D}\right)
=\frac 1D \left( 2+ \frac {l^2}{k^2}+\frac{(p-l)^2}{(p-k)^2} \right. 
\nonumber \\
&&\left. -\frac{(k-l)^2}{k^2} -\frac{(k-l)^2}{(p-k)^2}\right),
\eq
with
\be
D=k^2 l^2 (k-l)^2 (p-k)^2 (p-l)^2.
\ee
We disregard the surface term on momentum routing invariance grounds to 
write
\be
A_1=\int_{k,l}^{\Lambda}
\frac{1}{k^2 (p-k)^4 l^2 (l-k)^2}
\ee
and
\be
A_2=\int_{k,l}^{\Lambda}
\frac{1}{k^4 (p-k)^2 l^2 (p-l)^2}.
\ee
In order to reduce to these two integrals, we have performed shifts 
(surface terms
are not considered). The integral $A_2$ is actually a product of two 
independent
one-loop integrals in the internal momenta $l$ and $k$. It gives us
\bq
&&A_2= -\frac{p^2}{3} \left\{  I_{log}^2(\lambda^2)
-\frac {14}{3}b I_{log}(\lambda^2) \right. \nonumber \\
&& \left. +2b\ln{\left(-\frac {p^2}{\lambda^2}\right)}
I_{log}(\lambda^2)  \right. \nonumber \\
&& \left. + b^2 \ln ^2{\left(-\frac {p^2}{\lambda^2}\right)}
- \frac{14}{3} b^2 \ln{\left(-\frac {p^2}{\lambda^2}\right)} 
+\frac{16}{3} b^2 \right\}.
\nonumber \\
&&
\eq
For the $A_1$, after the integral in $l$ is solved, we have
\bq
&&A_1=-\frac 13 \int_k^{\Lambda} \frac{1}{(p-k)^4}
\left\{I_{log}(\lambda^2) \right. \nonumber \\
&& \left. +b \ln{\left(-\frac {k^2}{\lambda^2}\right)} -\frac 83 
b\right\}
\nonumber \\
&&= -\frac 13 \left\{I_{log}(\lambda^2)-\frac 83 
b\right\}\int_k^{\Lambda} \frac{1}{(p-k)^4}
\nonumber \\
&& -\frac b3 \int_k^{\Lambda} \frac{1}{(p-k)^4}
\ln{\left(-\frac {k^2}{\lambda^2}\right)}. \nonumber \\
&&
\eq
Using  an infrared cutoff, $m^2$, the first term
above is proportional to $I_{quad}(m^2)$, plus  a surface term. 
Therefore it does not contribute in the massless limit if we call for momentum 
routing invariance. Hence
\bq
&& A_1= -\frac b3 \int_k^{\Lambda} \frac{1}{[(p-k)^2-m^2]^2}
\ln{\left(-\frac {(k^2-m^2)}{\lambda^2}\right)} \nonumber \\
&&= -\frac b3 \int_k^{\Lambda} \frac{1}{[(p-k)^2-m^2]}
\ln{\left(-\frac {(k^2-m^2)}{\lambda^2}\right)} \times \nonumber \\
&&\left\{ \frac{1}{(k^2-m^2)}- \frac{p^2-2p \cdot k}{(k^2-m^2)^2}
+ \frac{(p^2-2p \cdot k)^2}{(k^2-m^2)^3} \right. \nonumber \\
&&\left. -\frac{(p^2-2p \cdot k)^3}{(k^2-m^2)^3[(k-p)^2-m^2]}
\right\}.
\label{A1}
\eq
It is easy to check that the first term above is $I^{(2)}$. If we use 
identity
(\ref{ident}) in $I^{(2)}$, we find
\bq
&&\int_k^{\Lambda}  \frac{p^2-2p \cdot k}{(k^2-m^2)^2[(p-k)^2-m^2]}
\ln{\left(-\frac {(k^2-m^2)}{\lambda^2}\right)}  \nonumber \\
&&=-I^{(2)}+I_{quad}^{(2)}(m^2)
=-I^{(2)}
\eq
and
\bq
&&\int_k^{\Lambda}  \frac{(p^2-2p \cdot k)^2}{(k^2-m^2)^3[(p-k)^2-m^2]}
\ln{\left(-\frac {(k^2-m^2)}{\lambda^2}\right)} \nonumber \\
&&=I^{(2)}-I_{quad}^{(2)}(m^2)+p^2I_{log}^{(2)}(m^2)\nonumber \\
&&=I^{(2)}+p^2I_{log}^{(2)}(m^2).
\eq
The last term of eq. (\ref{A1}) is finite and can be calculated with 
the use of eq. (\ref{ident2}).
The final result of $A_1$ is
\be
A_1=\ \frac {2b}{9} p^2\left\{I_{log}(\lambda^2)+b
\ln{\left(-\frac {p^2}{\lambda^2}\right)} +\frac 73 b \right\},
\ee
and we have, for the overlapping diagram,
\bq
&&-i \Sigma_2^{(2)}(p^2)=i \frac{g^4 p^2}{6} \left\{ 
-I_{log}^2(\lambda^2)
\right. \nonumber \\
&& \left. +\frac {16}{3}b I_{log}(\lambda^2) -2b\ln{\left(-\frac 
{p^2}{\lambda^2}\right)}
I_{log}(\lambda^2)  \right. \nonumber \\
&& \left. -b^2\ln^2{\left(-\frac {p^2}{\lambda^2}\right)}
+\frac {16}{3} b^2 \ln{\left(-\frac {p^2}{\lambda^2}\right)} - 
\frac{34}{9} b^2 \right\}.
\nonumber \\
&&
\eq
The result above refers to the complete diagram, which includes the 
subdivergences
with non-local terms.
The diagrams, according to the forest formula, for the counterterms to 
be
added so as to cancel the subdivergences are given
in fig. (\ref{fig3}).

We have two equal contributions that furnish us
\bq
&&-i\Sigma_{CT\,\,sub}^{(2)}=-2ig^2 
I_{log}(\lambda^2)(-i\Sigma(p^2))\nonumber \\
&&=\frac {ig^4p^2}{3}\Bigg\{
I_{log}^2(\lambda^2)+b\ln{\left(-\frac{p^2}{\lambda^2}\right)}
I_{log}(\lambda^2) \nonumber \\
&&-\frac 83 b I_{log}(\lambda^2)\Bigg\}
\eq
After adding the above counterterm, we finally obtain
\bq
&& -i\bar \Sigma^{(2)}_2(p^2)=
\frac {ig^4p^2}{6}\Bigg\{
I_{log}^2(\lambda^2)+ \nonumber \\
&&  -  b^2\ln^2{\left(-\frac {p^2}{\lambda^2}\right)}
+\frac {16}{3} b^2 \ln{\left(-\frac {p^2}{\lambda^2}\right)} - 
\frac{34}{9} b^2
\Bigg\}
\nonumber \\
&& \equiv (1 - t_{(a)} - t_{(b)}) (-i \Sigma_2^{(2)} (p^2)),
\eq
where $t_{(a)}$ and $t_{(b)}$ are the subtraction operators 
corresponding
to the renormalization parts in fig. (\ref{fig3}).

As in the one loop calculation, the limit $m^2 \to 0$ was taken in the 
end and we made use of
a scale relation  given by eq.
(\ref{scale2l}).
The complete self-energy at the $g^4$ order is given by
\bq
&& -i\Sigma^{(2)}(p^2)=\frac{ig^4p^2}{6}\Bigg\{ I^2_{log}(\lambda^2)+
\nonumber \\ && +
\frac 13 bI^{(2)}_{log}(\lambda^2) - \frac {10}{9}
b I_{log}(\lambda^2)
\Bigg\}  \nonumber \\ && - i \tilde \Sigma^{(2)}(p^2),
\eq
where the tilde refers to the finite part.
We are now able to  calculate the anomalous dimension 
($\gamma$-function).
As usual, we define the renormalization constants, $Z_3$ and $Z_g$, so 
that
\be
\phi_0=Z_3^{1/2}\phi \,\,\,\, \mbox{and} \,\,\,\,  g_0=Z_g g,
\ee
and the Lagrangian is written as
\bq
&& {\cal L}= \frac 12 \partial^\mu \phi \partial_\mu \phi +
\frac {g}{3!} \phi^3 + \frac 12 (Z_3-1) \partial^\mu \phi \partial_\mu
\phi
+ \nonumber \\ && + (Z_3^{3/2}Z_g -1)\frac {g}{3!} \phi^3.
\eq
The expansion in terms of the coupling constant, $g$, is given by
\be
Z_3= 1+\alpha_1 g^2+ \alpha_2g^4 +{\cal O}(g^6)
\ee
and
\be
Z_g= 1+\rho_1g^2+ \rho_2g^4 +{\cal O}(g^6).
\ee
According to our calculations, we have
\bq
\alpha_1&=&-\frac i6 I_{log}(\lambda^2), \\
\alpha_2&=& \frac 16 \Bigg\{ -I^2_{log}(\lambda^2)-\frac 13 b
I^{(2)}_{log}(\lambda^2)
+\nonumber \\ &+& \frac {10}{9}b
I_{log}(\lambda^2) \Bigg\}
\eq
and
\be
Z_3^{3/2}Z_g=1-ig^2I_{log}(\lambda^2),
\ee
which gives us
\be
\rho_1= -\frac 34 i I_{log}(\lambda^2).
\ee
 The $\gamma$-function is defined as
\be
\gamma=\frac {\lambda}{2Z_3}\frac {\partial Z_3}{\partial \lambda}=
\frac {\dot Z_3}{2 Z_3},
\ee
where
\be
\dot f=\lambda \frac{\partial f}{\partial \lambda}=2 \lambda^2
\frac {\partial f}{\partial \lambda^2}.
\ee
So, we have that
\be
\gamma= \frac 12 \left\{ \dot \alpha_1g^2 + (2 \beta_1 \alpha_1
+ \dot \alpha_2 - \alpha_1\dot \alpha_1)g^4 \right\}+ {\cal O}(g^6),
\nonumber
\ee
with $\beta_1$ the coefficient of $g^3$ in the $\beta$-function,
defined as
\be
\beta=\dot g= -g\frac {\dot Z_g}{Z_g}.
\ee
Taking into account that
\bq
\frac {d}{d\lambda^2} (I^2_{log}(\lambda^2))&=&\frac
{2b}{\lambda^2}I_{log}(\lambda^2), \\
\frac {d}{d\lambda^2} (I^{(2)}_{log}(\lambda^2))&=& \frac
{1}{\lambda^2}
\left( \frac 32 b - I_{log}(\lambda^2) \right)
\eq
and
\be
\frac {d}{d\lambda^2} (I_{log}(\lambda^2))= \frac {b}{\lambda^2},
\ee
we find
\be
\beta_1=-\frac {3}{4(4\pi)^3}
\ee
and
\be
\gamma=\frac {1}{12(4\pi)^3}g^2-\frac
{11}{108}
\frac{1}{4(4\pi)^6}g^4 + {\cal O}(g^6).
\ee
Notice that   $\beta_1$ and the $g^2$
contribution
to the $\gamma$-function are universal results whereas the $g^4$ 
coefficient
of the $\gamma$-function is renormalization scheme dependent. These 
results corresponds to the minimal subtration scheme in implicit 
regularization.

\section{Conclusions}
\indent

In this paper we have addressed important issues regarding the extension of implicit regularization beyond one loop order. Namely we showed that we can display the divergencies as basic 
loop integrals in one internal momentum in any loop order including the case of overlapping divergencies.
This is the heart of IR which enables us to work in the physical dimension of the model. Moreover we have shown the general form of surface terms which appear as finite differences of divergent integrals with the same superficial degree of divergence. They are important because they appear to relate momentum routing invariance to gauge and supersymmetry in a $n$-loop amplitude. We have explicitly verified this link in one loop order for both gauge  and supersymmetry and to three loop order for supersymmetry. A general proof of the connection between momentum routing invariance in a general Feynman diagram and the vital symmetries to be respected by the corresponding  Green's function is interesting to be constructed \cite{EDSON}. Besides
 We have also shown that  basic $n$-loop integrals may be 
written as  a function of an arbitrary parameter $\lambda$ which plays 
the role of renormalization group scale; they are absorbed as they stand  in the definition of the renormalization constants  
 in the light of the BPHZ forest formula in a regularization independent fashion. That is  because  
 the derivatives of the basic divergent integrals with respect 
to $\lambda$ can  be displayed by  loop integrals  as well.  
Finally, contrarily to dimensional methods where ultraviolet and infrared 
divergences may become mixed and lead to ambiguities, IR clearly separates ultraviolet and infrared degrees of freedom through a kind of scale relation obbeyed by the basic divergent integrals.

\end{document}